\documentclass[prl,twocolumn,a4paper,superscriptaddress]{revtex4}
\usepackage{graphicx,amsmath,bm,natbib, dsfont, longtable, amssymb}
\usepackage[bbgreekl]{mathbbol}

\def\bra#1{\mathinner{\langle{#1}|}}
\def\ket#1{\mathinner{|{#1}\rangle}}

\def\prjct#1{\mathinner{|{#1}\rangle}\!\!\mathinner{\langle{#1}|}}

\def\text#1{\textrm{#1}}
\def\id{\mathbb{1}}
\def\tr{\text{tr}}

\begin{document}

\title{What does it take to see entanglement?}
\author{Valentina Caprara Vivoli}
\affiliation{Group of Applied Physics, University of Geneva, CH-1211 Geneva 4, Switzerland}
\author{Pavel Sekatski}
\affiliation{Institut for Theoretische Physik, Universitat of Innsbruck, Technikerstra{\ss}e 25, A-6020 Innsbruck, Austria}
\author{Nicolas Sangouard}
\affiliation{Department of Physics, University of Basel, Klingelbergstrasse 82, 4056 Basel, Switzerland}

\date{\today}

\begin{abstract}
Tremendous progress has been realized in quantum optics for engineering and detecting the quantum properties of light. Today, photon pairs are routinely created in entangled states. Entanglement is revealed using single-photon detectors in which a single photon triggers an avalanche current. The resulting signal is then processed and stored in a computer. Here, we propose an approach to get rid of all the electronic devices between the photons and the experimentalist i.e. to use the experimentalist's eye to detect entanglement. We show in particular, that the micro entanglement that is produced by sending a single photon into a beam-splitter can be detected with the eye using the magnifying glass of a displacement in phase space. The feasibility study convincingly demonstrates the possibility to realize the first experiment where entanglement is observed with the eye.
\end{abstract}

\maketitle

\paragraph{Introduction ---}
The human eye has been widely characterized in the weak light regime. The data presented in Fig. 1 (circles) for example is the result of a well established experiment \cite{Hecht42} where an observer was presented with a series of coherent light pulses and asked to report when the pulse is seen (the data have been taken from Ref. \cite{Rieke98}). While rod cells are sensitive to single photons \cite{Phan14}, these results show unambiguously that one needs to have coherent states with a few hundred photons on average, incident on the eye to systematically see light. As mentioned in Ref. \cite{Sekatski09}, the results of this experiment are very well reproduced by a threshold detector preceded by loss. In particular, the red dashed line has been obtained with a threshold at 7 photons combined with a beamsplitter with 8\% transmission efficiency. In the low photon number regime, the vision can thus be described by a positive-operator valued measure (POVM) with two elements $P_{\text{ns}}^{\theta,\eta}$ for \textquotedblleft not seen\textquotedblright and $P_{\text{s}}^{\theta,\eta}$ for \textquotedblleft seen\textquotedblright where $\theta=7$ stands for the threshold, $\eta=0.08$ for the efficiency, see Appendix, part I. It is interesting to ask what it takes to detect entanglement with such a detector. \\

\begin{figure}[ht!]
\includegraphics[width=0.45\textwidth]{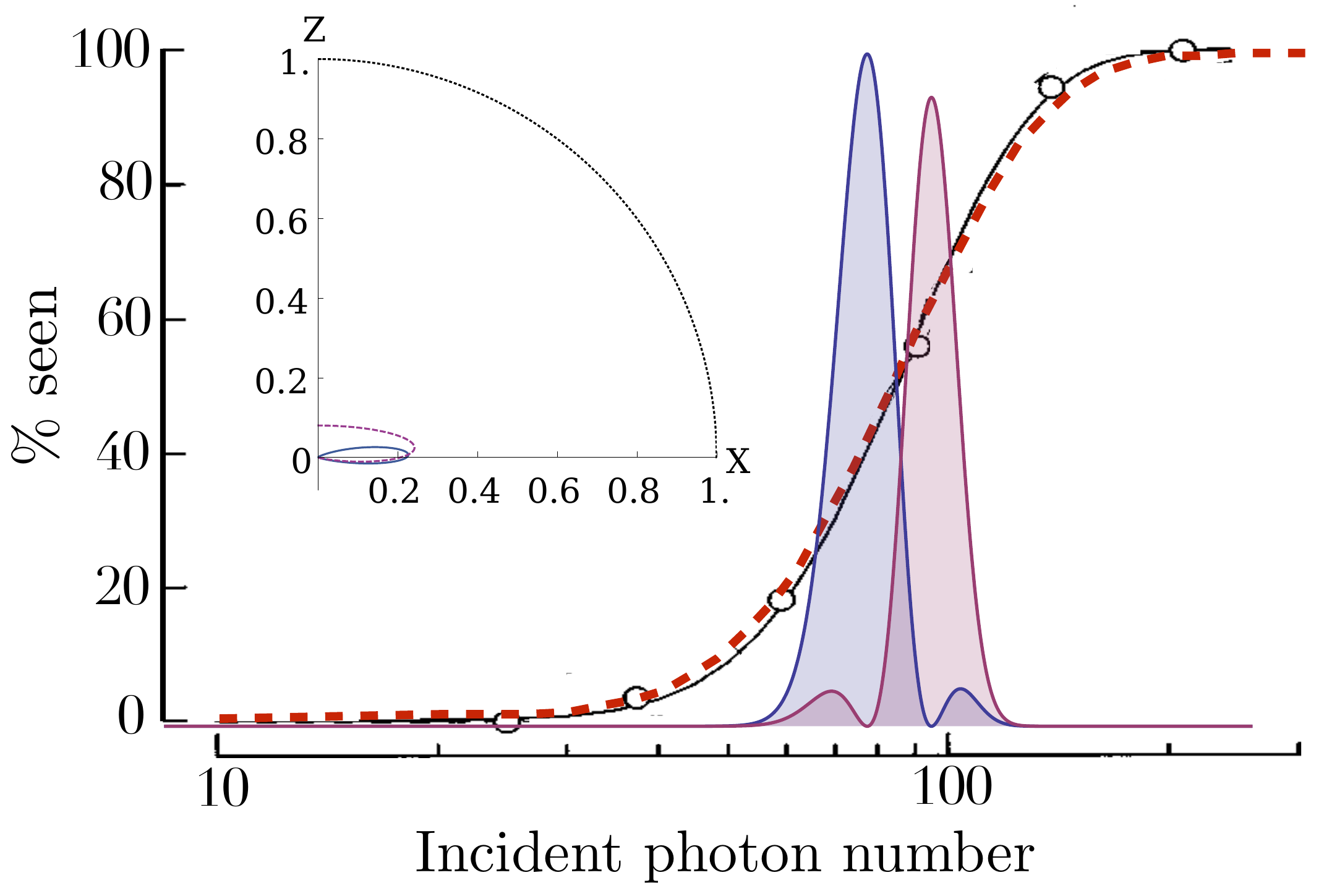}
\caption{Experimental results (circles) showing the probability to see coherent light pulses as a function of the mean photon number (taken from Ref. \cite{Rieke98}). The black line is a guide for the eye. The dashed red line is the response of a threshold detector with loss (threshold at 7 photons and 8\% efficiency). Such a detector can be used to distinguish the states $\ket{0+1}$ and $\ket{0-1}$ when they are displaced in phase space: The displacement operation not only increases the photon number but also makes the photon distribution distinguishable. This is shown through the two bumps which are the photon number distribution of $\ket{D(\alpha)(0+1)}$ and $\ket{D(\alpha)(0-1)}$ respectively for $\alpha \sim \sqrt{100}$. The inset is a quarter of the xz plane of the Bloch sphere having the vacuum and single photon Fock states $\{|0\rangle, |1\rangle \}$ as north and south poles respectively. A perfect qubit measurement corresponds to a projection along a vector with unit length (dotted line). The POVM element \textquotedblleft no click\textquotedblright of a measurement combining a single-photon detector with 8\% efficiency and a displacement operation defines a non-unit vector on the sphere for which the angle with the z axis can be changed by tuning the amplitude of the displacement (purple dashed curve). For a displacement with a zero amplitude (no displacement), this vector points out in the z direction whereas for an amplitude $\sim \sqrt{12.5},$ the vector points out in the x direction. The POVM element \textquotedblleft not seen\textquotedblright of a measurement combining a human eye with a displacement operation also defines a non-unit vector on the sphere. The angle between this vector and the z axis can also be varied by changing the size of the displacement. In particular, for an amplitude of the displacement of $\sim \sqrt{100}$, this vector points out in the x direction and in this case, the measurement with the eye is fairly similar to the measurement with the single-photon detector with the same efficiency. Rotation in the xy plane can be obtained by changing the phase of the displacement operation.}
\label{Fig1}
\end{figure}

Let us note first that such detection characteristics do not prevent the violation of a Bell inequality. In any Bell test, non-local correlations are ultimately revealed by the eye of the experimentalist, be it by analyzing numbers on the screen of a computer or laser light indicating the results of a photon detection. The subtle point is whether the amplification of the signal prior to the eye is reversible. Consider a gedanken experiment where a polarization-entangled two photon state $\frac{1}{\sqrt{2}}(\ket{h}_A \ket{v}_B-\ket{v}_A \ket{h}_B)$ is shared by two protagonists -- Alice and Bob -- who  easily rotate the polarization of their photons with wave plates. Assume that they can amplify the photon number with the help of some unitary transformation $U$ mapping, say, a single photon to a thousand photons while leaving the vacuum unchanged. It is clear that in this case, Alice and Bob can obtain a substantial violation of the Bell-CHSH inequality \cite{CHSH69}, as the human eye can almost perfectly distinguish a thousand photons from the vacuum. In practice, however, there is no way to properly implement $U.$ Usually, the amplification is realized in an irreversible and entanglement-breaking manner, e.g. in a measure and prepare setting with a single photon detector triggering a laser \cite{Pomarico11}. In this case however the detection clearly happens before the eye. \\

One may then wonder whether there is a feasible way to reveal entanglement with the eye in reversible scenarios, i.e. with states, rotations and unitary amplifications that can be accessed experimentally. The task is a priori challenging. For example, the proposal of Ref. \cite{Brunner08} where many independent entangled photon pairs are observed does not allow one to violate a Bell inequality with the realistic model of the eye described before. A closer example is the proposal of Ref. \cite{Sekatski09} where entanglement of a photon pair is amplified through a phase covariant cloning. Entanglement can be revealed with the human eye in this scenario if strong assumptions are made on the source. For example, a separable model based on a measure and prepare scheme, has shown that it is necessary to assume that the source produces true single photons \cite{Sekatski10, Pomarico11}. Here, we go beyond such a proposal by showing that entanglement can be seen without assumption on the detected state. Inspired by a recent work  \cite{Monteiro15}, we show that it is possible to detect path entanglement, i.e. entanglement between two optical paths sharing a single photon, with a trusted model of the human eye upgraded by a displacement in phase space. The displacement operation which serves as a photon amplifier, can be implemented with an unbalanced beamsplitter and a coherent state \cite{Paris96}. Our proposal thus relies on simple ingredients. It does not need interferometric stabilization of optical paths and is very resistant to loss. It points towards the first experiment where entanglement is revealed with human eye-based detectors.\\

\paragraph{Upgrading the eye with displacement ---}
Our proposal starts with an entangled state between two optical modes $A$ and $B$
\begin{equation}
\label{path_ent}
|\psi_+\rangle=\frac{1}{\sqrt{2}}\left(|0\rangle_A|1\rangle_{B} + |1\rangle_A|0\rangle_B \right).
\end{equation}
Here $|0\rangle$ and $|1\rangle$ stands for number states filled with the vacuum and a single photon respectively.
To detect entanglement in state (\ref{path_ent}), a method using a photon detector -- which does not resolve the photon number $(\theta=1)$ --  preceded by a displacement operation has been proposed in Ref. \cite{Banaszek98} and used later in various experiments \cite{Kuzmich00, Hessmo04, Monteiro15}. In the $\{ |0\rangle,|1\rangle \}$ subspace, this measurement is a two outcome $\{P_{\text{ns}}^{1,\eta}$ for \textquotedblleft no click\textquotedblright, $P_{\text{s}}^{1,\eta}$ for \textquotedblleft click\textquotedblright $\}$ non-extremal POVM on the Bloch sphere whose direction depends on the amplitude and phase of the displacement \cite{Caprara15b}. In particular, pretty good measurements can be realized in the x direction. This can be understood by realizing that the photon number distribution of the two states $|\mathcal{D}(\alpha)(0+1)\rangle$ and $|\mathcal{D}(\alpha)(0-1)\rangle$ where $\mathcal{D}(\alpha)$ is the displacement, slightly overlap in the photon number space and their mean photon numbers differ by $2 |\alpha|,$ see Fig. \ref{Fig1}. This means that they can be distinguished, at least partially, with threshold detectors. It is thus interesting to analyze an eye upgraded by a displacement operation. In the $\{|0\rangle, |1\rangle\}$ subspace, we found that the elements $\{P_{\text{ns}}^{7,\eta}, P_{\text{s}}^{7,\eta}\}$ also constitute a non-extremal POVM, and as before, their direction in the Bloch sphere depends on the amplitude and phase of the displacement. For comparison, the elements \textquotedblleft no click\textquotedblright and \textquotedblleft not seen\textquotedblright are given in the inset of Fig. \ref{Fig1} considering real displacements and focusing on the case where the efficiency of the photon detector is equal to 8\%. While the eye-based measurement cannot perform a measurement in the z direction, it is comparable to the single photon detector for performing measurements along the x direction. Identical results would be obtained in the yz plane for purely imaginary displacements. More generally, the measurement direction can be chosen in the xy plane by changing the phase of the displacement. We present in the next paragraph an entanglement witness suited for such measurements.\\

\begin{figure}
\includegraphics[width=0.45\textwidth]{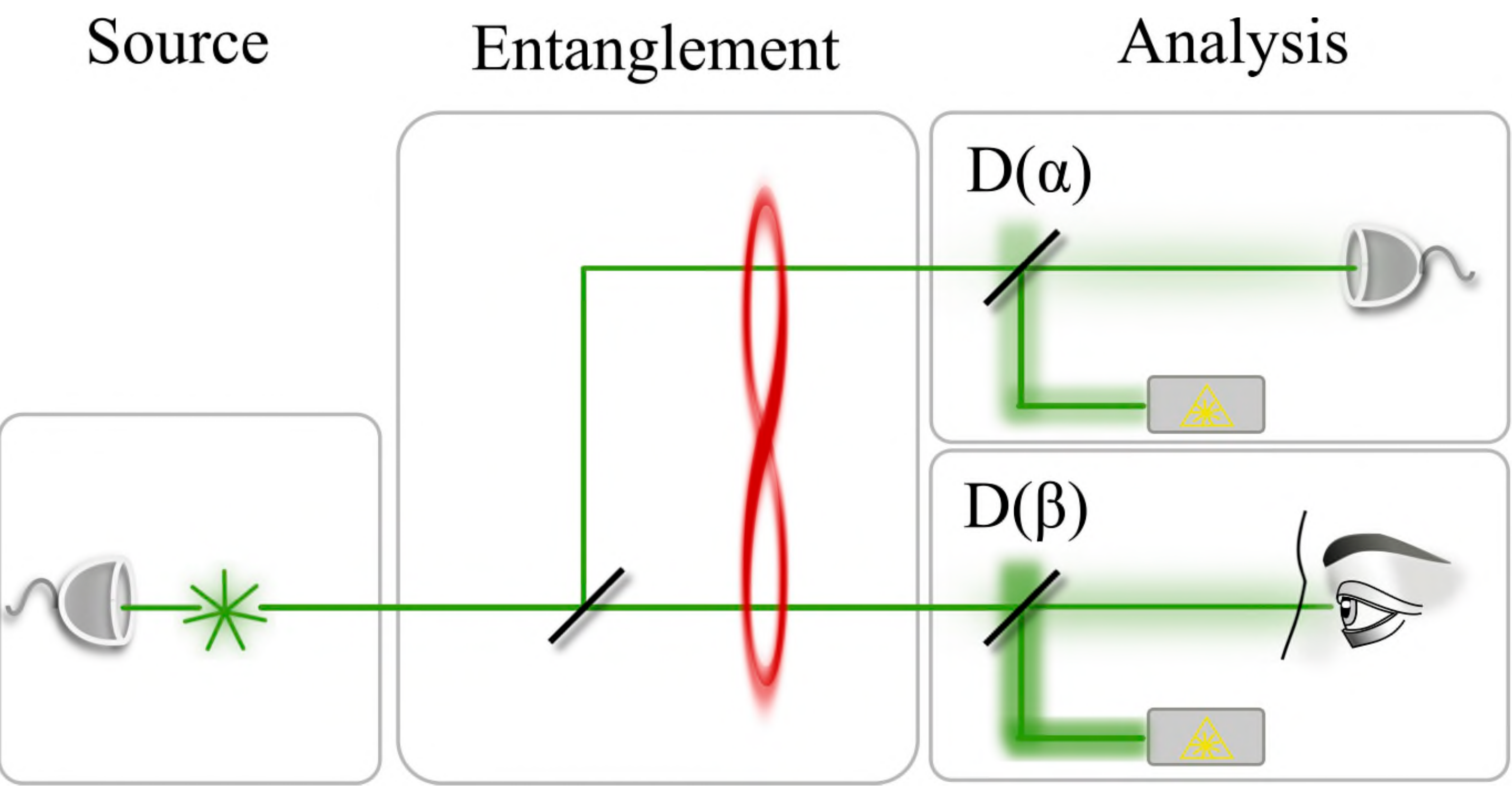}
\caption{Scheme of our proposal for detecting entanglement with the human eye. A photon pair source based on spontaneous parametric down conversion is used as a single photon source, the emission of a photon being heralded by the detection of its twin. The heralded photon is then sent into a beamsplitter to create path entanglement, i.e. entanglement between two optical modes sharing a delocalized single photon. The entangled state is subsequently detected using a photon counting detector preceded by a displacement operation on one mode, and using a human eye preceded by a displacement on the other mode. The correlations between the results (click and no click for the photon detector, seen and not seen for the human eye) allows one to conclude about the presence of entanglement, c.f. main text}
\label{Fig2}
\end{figure}

\paragraph{Witnessing entanglement with the eye ---}
We consider a scenario where path entanglement is revealed with displacement operations combined with a photon detector on mode A and with the eye on mode B, c.f. Fig. \ref{Fig2}. We focus on the following witness
\begin{equation}
W=\int_0^{2\pi} \frac{d\varphi}{2\pi} U_{\varphi}^{\dag} \otimes U_{\varphi}^{\dag} \left(\bbsigma_{\alpha}^{1} \otimes \bbsigma_{\beta}^{7} \right) U_{\varphi} \otimes U_{\varphi}
\end{equation}
where $\bbsigma_{\alpha}^{\theta}=D(\alpha)^\dag \left(2P_{\text{ns}}^{\theta,1}-\mathbb{1}\right)D(\alpha)$ is the observable obtained by attributing the value +$1$ to events corresponding to \textquotedblleft no click\textquotedblright (\textquotedblleft not seen\textquotedblright) and -$1$ to those associated to \textquotedblleft click\textquotedblright (\textquotedblleft seen\textquotedblright). Since we are interested in revealing entanglement at the level of the detection, the inefficiency of the detector can be seen as a loss operating on the state, i.e. the beamsplitter modeling the detector inefficiency acts before the displacement operation whose amplitude is changed accordingly \cite{Monteiro15}. This greatly simplifies the derivation of the entanglement witness as this allows us to deal with detectors with unit efficiencies ($\eta=1$). The phase of both displacements $\alpha$ and $\beta$ is randomized through the unitary transformation $U_\varphi=e^{i \varphi a^{\dag}a}$ for A (where $a,$ $a^\dag$ are the bosonic operators for the mode A) and similarly for B. The basic idea behind the witness can be understood by noting that for ideal measurements $W_{\text{ideal}}=\int (\cos \varphi \sigma_x + \sin \varphi  \sigma_y)  \!\otimes\! (\cos \varphi  \sigma_x + \sin \varphi  \sigma_y) \frac{d\varphi}{2\pi}$ equals the sum of coherence terms $|01\rangle \langle 10 | + |10 \rangle \langle 01|.$ Since two qubit separable states stay positive under partial transposition \cite{Peres96, Horodecki96}, these coherence terms are bounded by $2 \sqrt{p_{00} p_{11}}$ for two qubit separable states where $p_{ij}$ is the joint probability for having $i$ photons in A and $j$ photons in B. Any state $\rho$ such that $\tr \big[\rho W_{\text{ideal}} \big] >  2 \sqrt{p_{00} p_{11}}$ is thus necessarily entangled. Following a similar procedure, we find that for any two qubit separable states, $\tr \big[W \rho_{\text{sep}}^{\text{qubit}}\big] \leq W_{\text{ppt}}$ where
\begin{equation}
W_{\text{ppt}} =  \sum_{i,j = 0}^{1} \langle ij| W |ij\rangle p_{ij} +  2 |\langle 10| W |01\rangle| \sqrt{p_{00}p_{11}},
\end{equation}
see Appendix, part II. The $p_{ij}$s can be bounded by noting that for well chosen displacement amplitudes, different photon number states lead to different probabilities \textquotedblleft not seen\textquotedblright and \textquotedblleft no click\textquotedblright. For example, we show in the Appendix, part III that
\begin{equation}
\nonumber
p_{00}\le \frac{P_{AB}(\text{+}1\text{+}1|0 \beta_0,\rho_{\text{exp}}) - P_B(\text{+}1|\beta_0,\ket{1})P_{A}(\text{+}1|0,\rho_{\text{exp}})}{P_B(\text{+}1|\beta_0,\ket{0})-P_B(\text{+}1|\beta_0,\ket{1})}.
\end{equation}
$P_B(+1 | \beta_0, \rho_{\text{exp}})$ is the probability \textquotedblleft not seen\textquotedblright  when looking at the experimental state $\rho_{\text{exp}}$ amplified by the displacement $\beta_0.$ This is a quantity that is measured, unlike $P_B(+1 | \beta_0, |1\rangle),$ which is computed from  $\frac{1}{2}\left(1 + \langle 1 | \sigma_{\beta_0}^7 |1\rangle\right).$ $\beta_0$ is the amplitude of the displacement such that $P_B(+1 | \beta_0, |0\rangle)=P_B(+1 | \beta_0, |2\rangle).$ $p_{10},$ $p_{11}$ and  $p_{01}$ can be bounded in a similar way, the two latter requiring another displacement amplitude $\beta_1$, see Appendix, part III.\\

The recipe that we propose for testing the capability of the eye to see entanglement thus consists in four steps. i) Measure the probability that the photon detector in A does not click and of the event \textquotedblleft not seen\textquotedblright for two different displacement amplitudes  $\{0, \beta_0\},$ $\{0,\beta_1\}.$ ii) Upper bound from i) the joint probabilities $p_{00},$ $p_{11}$ $p_{01}$ and $p_{10}.$ iii) Deduce the maximum value that the witness W would take on separable states $W_{\text{ppt}}$. iv) Measure $\langle W \rangle.$ If there are values of $\alpha$ and $\beta$ such that $\langle W \rangle > W_{\text{ppt}},$ we can conclude that the state is entangled. Note that this conclusion holds if the measurement devices are well characterized, i.e. the models that are used for the detections well reproduce the behavior of single photon detectors and eyes, the displacements are well controlled operations and filtering processes ensure that a single mode of the electromagnetic wave is detected. We have also assumed hitherto that the measured state is well described by two qubits. In the Appendix part IV, we show how to relax this assumption by bounding the contribution from higher photon numbers. We end up with an entanglement witness that is state independent, i.e. valid independently of the dimension of the underlying Hilbert space.\\

\begin{figure}
\includegraphics[width=0.45\textwidth]{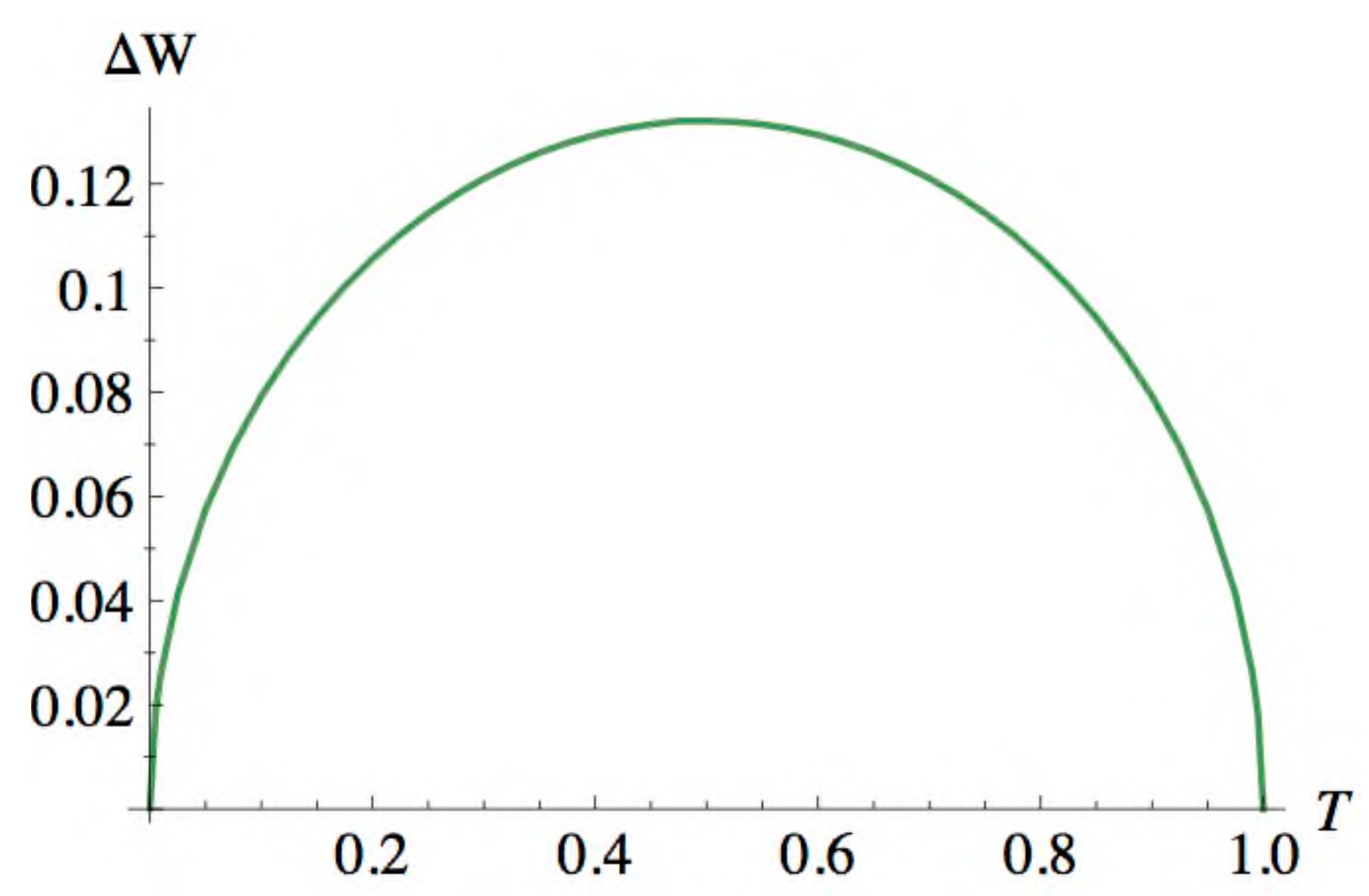}
\caption{Value of the witness that would be measured in the setup shown in Fig. \ref{Fig2} relative to the value that would be obtained from state with a positive partial transpose $\Delta W = \langle W \rangle - W_{\text{ppt}}$ as a function of the beamsplitter transmission efficiency $T$ under realistic assumption about efficiencies, c.f. main text.}
\label{Fig3}
\end{figure}

\paragraph{Proposed setup ---}
The experiment that we envision is represented in Fig. \ref{Fig2}. A single photon is generated from a photon pair source and its creation is heralded through the detection of its twin photon. Single photons at 532 nm can be created in this way by means of spontaneous parametric down conversion \cite{Phan14}. They can be created in pure states by appropriate filtering of the heralding photon, see e.g. \cite{Monteiro15}. The heralded photon is then sent into a beamsplitter (with transmission efficiency $T$) which leads to entanglement between the two output modes. As described before, displacement operations upgrade the photon detection in A and the experimentalist's eye in B. In practice, the local oscillators needed for the displacement can be made indistinguishable from single photons by using a similar non-linear crystal pumped by the same laser but seeded by a coherent state, see e.g. \cite{Bruno13}. The relative value $\Delta W = \langle W \rangle - W_{\text{ppt}}$ that would be obtained in such an experiment is given in Fig. \ref{Fig3} as a function of $T.$ We have assumed a transmission efficiency from the source to the detectors $\eta_t = 90\%,$ a detector efficiency of $80\%$ in A and an eye with the properties presented before (8\% efficiency and a threshold at 7 photons). The results are optimized over the squeezing parameter of the pair source for suitable amplitudes of the displacement operations, see Appendix, part V. We clearly see that despite low overall efficiencies and multi-photon events that are unavoidable in spontaneous parametric down conversion processes, our entanglement witness can be used to successfully detect entanglement with the eye. Importantly, there is no stabilization issue if the local oscillator that is necessary for the displacement operations is superposed to each mode using a polarization beamsplitter instead of a beamsplitter to create path entanglement, see e.g. \cite{Morin13}. The main challenge is likely the timescale of such an experiment, as the repetition time is inherently limited by the response of the experimentalist, but this might be overcome, at least partially by measuring directly the response of rod cells as in Ref. \cite{Phan14}. \\

\paragraph{Conclusion ---}
Our results help in clarifying the requirements to see entanglement. If entanglement breaking operations are used, as in the experiments performed so far, it is straightforward to see entanglement. In this case, however, the measurement happens before the eyes. In principle, the experimentalist can reveal non-locality directly with the eyes from reversible amplifications, but these unitaries cannot be implemented in practice. What we have shown is that entanglement can be realistically detected with human eyes upgraded by displacement operations in a state-independent way. From a conceptual point of view, it is interesting to wonder whether such experiments can be used to test collapse models in perceptual processes in the spirit of what has been proposed in Refs.  \cite{Ghirardi99,Thaheld03}. For more applied perpectives, our proposal shows how threshold detectors can be upgraded with a coherent amplification up to the point where they become useful for quantum optics experiments. Anyway, it is safe to say that probing the human vision with quantum light is a \textit{terra incognita}. This makes it an attractive challenge on its own. \\

\paragraph{Acknowledgements ---} We thank C. Brukner, W. D\"ur, F. Fr\"owis, N. Gisin, K. Hammerer, M. Ho, M. Munsch, R. Schmied, A. S{\o}rensen, P. Treutlein, R. Warburton and P. Zoller for discussions and/or comments on the manuscript. This work was supported by the Swiss National Science Foundation (SNSF) through NCCR QSIT and Grant number PP00P2-150579, by the John Templeton Foundation, and by the Austrian Science Fund (FWF), Grant number J3462 and P24273-N16.\\

\paragraph{Appendix I}
In this section, we provide a convenient expression for a threshold detector with non-unit efficiency (threshold $\theta$ and efficiency $\eta$). By modeling loss by a beamsplitter, the no-click event can be written as $P_{\text{ns}}^{\theta,\eta}=C_L^{\dagger}  \sum_{m=0}^{\theta -1}\ket{m}\bra{m} C_L$ where $C_L=e^{\tan \gamma\, a c^{\dagger}}e^{\ln(\cos\gamma)a^{\dagger}a}\ket{0}_c$ stands for the beam splitter. $a,$ $a^\dag$ are the bosonic operators for the detected mode and $\cos^2 \gamma=\eta$. After straightforward manipulations we can find that
\begin{equation}\label{Pns}
 P_{\text{ns}}^{\theta,\eta}= \frac{\eta^\theta}{(\theta-1)!} \frac{d^{\theta-1}}{d(1-\eta)^{\theta-1}}\frac{(1-\eta)^{a^\dag a}}{\eta}.
\end{equation}
The click event can be deduced from $P_{\text{s}}^{\theta,\eta}= \id -P_{\text{ns}}^{\theta,\eta}$.\\

\begin{figure}[ht]
\begin{center}
\includegraphics[width=230pt]{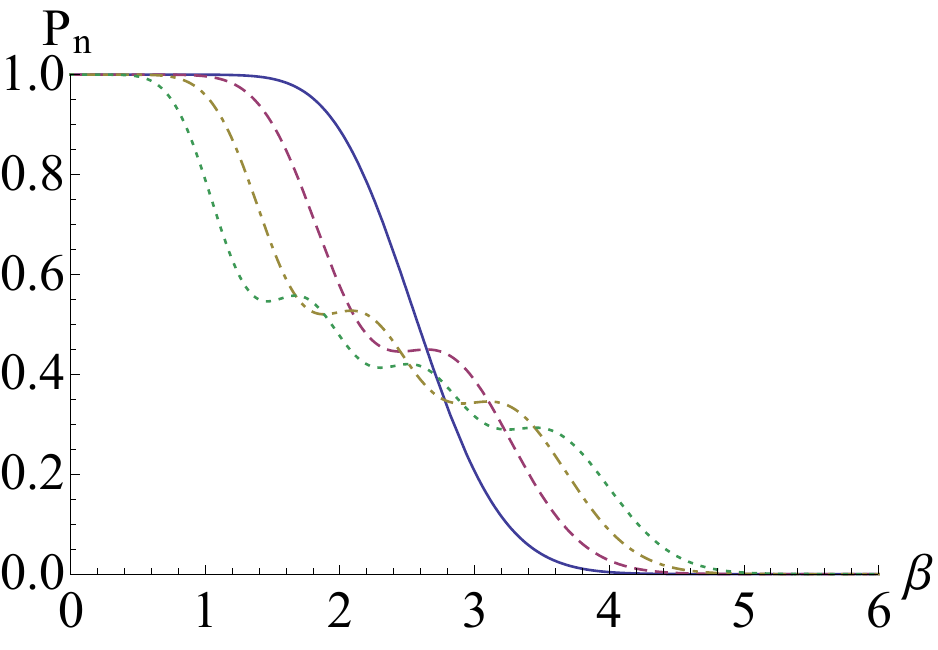}
\end{center}
\caption{Probability for having no click on a threshold detector $(\theta=7)$ with a number state $\ket{n}$ that is displaced in phase space as a function of the displacement amplitude $\beta$\label{Pn}}
\end{figure}

\paragraph{Appendix II}
Here we give details on how the entanglement witness has been derived, assuming first that one has qubits. Let's consider a general density matrix $P$ in the subspace $\{\ket{0},\ket{1}\}$. We look for the maximal value that $\langle W \rangle $ can take over the states staying positive under partial transposition, i.e. we want to optimize $\langle W \rangle $ over  $P$ such that i) $P\ge 0$, ii) $\tr(P)= 1$ and iii) $P^{T_b}\ge 0$. Here $P^{T_b}$ stands for the partial transposition over one party. As $\langle W \rangle$ is non-zero in blocks spanned by $\{ \ket{00}\}$, $\{\ket{01},\ket{10}\}$ and $\{\ket{11}\}$ only, it is straightforward to show that for any separable state
\begin{equation}\label{WPPT01}
W_{\text{ppt}}=\sum_{i,j=0}^1 \bra{ij}W\ket{ij} p_{ij}+2 \, |\bra{01}W\ket{10}| \sqrt{p_{00} p_{11}},
\end{equation}
where $p_{ij}=\bra{ij}P\ket{ij}.$ Any state $\rho_{\text{exp}}$ for which
$
\tr(\rho_{\text{exp}} \, W)- W_{\text{ppt}} > 0
$
has a negative partial transpose, i.e. is necessarily entangled.  It is important to stress that $W_{\text{ppt}}$ depends on the photon number statistics $\vec p = p_{ij}.$ We show in the next section how they can bounded. \\

\paragraph{Appendix III}
Figure \ref{Pn} shows the probability for having no click on a threshold detector $(\theta=7)$ with a number state $\ket{n}$ that is displaced in phase space as a function of the  displacement amplitude $\beta$, $P_B(+1|\beta,\ket{n})$ for $n=0$, $1$, $2$, $3.$  We show how to bound $p_{00}$, $p_{01}$, $p_{10}$, and $p_{11}$ from these results.\\

In order to bound $p_{00}$ and $p_{01},$ let's consider the displacement amplitude $\beta_0$ $(\sim 2.71)$ such that $P_B(+1|\beta_0,\ket{0})=P_B(+1|\beta_0,\ket{2}).$ We have

\begin{widetext}
\begin{eqnarray}
\nonumber
&P_{AB}(+1+1|0 \beta_0,\rho_{\text{exp}})&=\sum_{n=0}^{+\infty} p_{0n} P_B(+1|\beta_0,\ket{n})\\
\nonumber
&&\le P_B(+1|\beta_0,\ket{1})p_{0A}+(P_B(+1|\beta_0,\ket{0})-P_B(+1|\beta_0,\ket{1}))p_{00}.
\end{eqnarray}
using $P_B(+1|\beta_0,\ket{n\ge 2})<P_B(+1|\beta_0,\ket{1}).$ Note that $p_{0n}=\bra{0n}\rho_{\text{exp}}\ket{0n}$ and $p_{0A} = \tr (\rho_{\text{exp}} \ket{0}_A \bra{0}).$ This leads to the upperbound
\begin{equation}
\nonumber
p_{00}\le \frac{P_{AB}(+1+1|0 \beta_0,\rho_{\text{exp}})-P_B(+1|\beta_0,\ket{1})P_{A}(+1|0,\rho_{\text{exp}})}{P_B(+1|\beta_0,\ket{0})-P_B(+1|\beta_0,\ket{1})}.
\end{equation}
In the same way, we get
\begin{equation}
\nonumber
p_{01}\le \frac{P_{AB}(-1+1|0 \beta_0,\rho_{\text{exp}})-P_B(+1|\beta_0,\ket{1})P_{A}(-1|0,\rho_{\text{exp}})}{P_B(+1|\beta_0,\ket{0})-P_B(+1|\beta_0,\ket{1})}.
\end{equation}

To bound $p_{10}$ and $p_{11}$ we consider the displacement amplitude  $\beta_1$ $(\sim 2.09)$ such that $P_B(+1|\beta_1,\ket{1})=P_B(+1|\beta_1,\ket{2})$ (and $P_B(+1|\beta_1,\ket{n\ge 3})\le P_B(+1|\beta_1,\ket{1})$.) We get
\begin{equation}
\nonumber
p_{10}\le \frac{P_{AB}(+1+1|0 \beta_1,\rho_{\text{exp}})-P_B(+1|\beta_1,\ket{0})P_{A}(+1|0,\rho_{\text{exp}})}{P_B(+1|\beta_1,\ket{1})-P_B(+1|\beta_1,\ket{0})}.
\end{equation}
\begin{equation}
\nonumber
p_{11}\le \frac{P_{AB}(-1+1|0 \beta_1,\rho_{\text{exp}})-P_B(+1|\beta_1,\ket{0})P_{A}(-1|0,\rho_{\text{exp}})}{P_B(+1|\beta_1,\ket{1})-P_B(+1|\beta_1,\ket{0})}.
\end{equation}

Note also that for $\beta_2$ ($\sim 2.64$) such that $P_B(+1|\beta_2,\ket{0})=P_B(+1|\beta_2,\ket{1})$ (and $P_B(+1|\beta_2,\ket{n\ge 2})<P_B(+1|\beta_2,\ket{0})$), we have
\begin{eqnarray}
\nonumber
&& p_{n\ge2B} = \sum_{n \ge 2} \tr( \rho_{\text{exp}} \ket{n}\bra{n}_B )  \le \frac{P_B(+1|\beta_2,\rho_{\text{exp}})-P_B(+1|\beta_2,\ket{0})}{P_B(+1|\beta_2,\ket{3})-P_B(+1|\beta_2,\ket{0})}=p_B^*.
\end{eqnarray}
\end{widetext}
Note that $p_{n\ge2A}$ can be bounded from an auto-correlation measurement (see Ref. [13] of the main text). The upperbound on $p_{n\ge2A}$ is called $p_A^*.$
Importantly, the previous upperbounds hold in the qudit case, i.e. if the modes A and B are filled with more than one photon. \\

\paragraph{Appendix IV}
Now consider the case where the state has an arbitrary dimension in the Fock space. We can proceed as follows. A generic state $P$ can be written as
\begin{equation}
P=\left(
\begin{array}{cc}
P_{n_a\le1\cap n_b\le1}&P_{\text{coh}}\\
P_{\text{coh}}^{\dagger}&P_{n_a\ge2\cup n_b\ge2}\\
\end{array}
\right).
\end{equation}
We focus on the detection of entanglement in the qubit subspace $P_{n_a\le1\cap n_b\le1}.$ By linearity of the trace, we have
\begin{align}
\tr(W P) &= \tr(P_{n_a\le1\cap n_b\le1} W) +\tr\Big((P_{coh}^\dag + P_{coh})W\Big) \nonumber\\&+ \tr(P_{n_a\ge2\cup n_b\ge2} W).
\end{align}
Let us treat those terms one by one.
The maximum algebraic value of W is equal to 1, in such a way that the third term is upperbounded by $\tr(P_{n_a\ge2\cup n_b\ge2} W)\leq \tr(P_{n_a\ge2\cup n_b\ge2}) \leq p_A^*+p_B^*=p^*.$

The first term  is the subject of the second section, where we showed that $\tr(W P_{n_a\le1\cap n_b\le1}) \leq W_\text{ppt}(\vec p)$ given in \eqref{WPPT01}.

To bound the second term, let us recall that $W$ does not contain coherences between sectors of different total photon number, in such a way that
\begin{equation}
\nonumber
\tr\Big((P_{coh}^\dag + P_{coh})W\Big) \leq 2( |C_{11}^{20} W_{11}^{20}|+ |C_{11}^{02} W_{11}^{02}|),
\end{equation}
where $C_{ij}^{kl} = \bra{ij} P \ket{kl}$ and $W_{ij}^{kl} = \bra{ij} W \ket{kl}$. The positivity of the state $P$ restricted to the subspace $\{\ket{20}, \ket{02}, \ket{11}\}$ implies $C_{11}^{kl}\le \sqrt{p_{11}p_{kl}}$. Since $p_{20} \leq p_{A}^*$ and $p_{02} \leq p_{B}^*,$ we have
\begin{equation}
\nonumber
\tr\Big((P_{coh}^\dag + P_{coh})W\Big) \leq 2 \sqrt{p_{11}}\left(|W_{11}^{20}|\sqrt{p_A^*}+|W_{11}^{02}|\sqrt{p_{B}^*}\right),
\end{equation}

Finally, any state $P$, such that its restriction $P_{n_a\le1\cap n_b\le1}$ remains positive under partial transpose, satisfies
\begin{align}
\tr(W P) &\leq W_{\text{PPT}} \nonumber\\
\nonumber
&= W_{\text{ppt}}(\vec p)+ 2 \sqrt{p_{11}}\left(|W_{11}^{20}|\sqrt{p_A^*}+|W_{11}^{02}|\sqrt{p_{B}^*}\right) \\
\nonumber
& \-\ \-\  + p^*.
\end{align}
Any state $\rho_{\text{exp}}$ such that $\tr (W \rho_{\text{exp}}) - W_{\text{PPT}} >0$ is necessary entangled.\\

\paragraph{Appendix V}
The value of W that would be observed in the experiment represented in Fig. 2 of the main text can be calculated from
\begin{equation}
\nonumber
\langle W\rangle=\tr\left(\bbsigma^{7,\eta_b}_{\beta} \bbsigma^{1,\eta_a}_{\alpha}\rho_h\right),
\end{equation}
where $\eta_a$ and $\eta_b$ are the efficiencies of the detector in mode A and B respectively. $\rho_h$ is the density matrix after the beamsplitter (with transmission $T$) that is conditioned on a click in the heralding detector. The amplitude of the displacements are chosen such that $\beta=\sqrt{\frac{7}{\eta_b}}$, $\alpha=\frac{1}{\sqrt{\eta_a}}.$ Given the efficiency of the heralding detector $\eta_h=1-R_h$ and the squeezing parameter $g$ of the SPDC source, the state that is announced by a click on the heralding detector can be expressed as a difference of two
thermal states
\begin{widetext}
\nonumber
\begin{equation}
\frac{1-R_h^2T^2_{g}}{T^2_{g}\left(1-R_h^2\right)}\Bigg[\rho_{\text{th}}\left(\bar{n}=\frac{T_g^2}{1-T_g^2}\right) -\frac{1-T_g^2}{1-R_h^2T_g^2}\rho_{\text{th}}\left(\bar{n}=\frac{R_h^2T_g^2}{1-R_h^2T_g^2}\right)\Bigg]
\end{equation}
where $T_g=\tanh g$ and $\rho_{\text{th}}(\bar{n})=\frac{1}{1+\bar{n}}\sum_{k}\left(\frac{\bar{n}}{1+\bar{n}}\right)^k\ket{k}\bra{k}.$ We get
\begin{align}
\nonumber
\langle W\rangle &=\frac{1-R_h^2T^2_{g}}{T^2_{g}\left(1-R_h^2\right)}\Bigg[W^{\text{th}}\left(\bar{n}=\frac{T_g^2}{1-T_g^2}\right)
-\frac{1-T_g^2}{1-R_h^2T_g^2}W^{\text{th}}\left(\bar{n}=\frac{R_h^2T_g^2}{1-R_h^2T_g^2}\right)\Bigg]
\end{align}
where
\begin{align}
\nonumber
W^{\text{th}}(\bar n)=\frac{\eta_b^7}{6!} \frac{d^6}{d(1-\eta_b)^{6}}\frac{1}{\eta_b}\Bigg[&1+4\frac{ e^{-\eta_a|\alpha| ^2 -\eta_b|\beta| ^2 +\frac{\bar{n} \left|\alpha  \eta_a \sqrt{R}+\beta  \eta_b\sqrt{T} \right|^2}{\bar{n} (\eta_a R+T \eta_b)+1}}}{\bar{n} (\eta_a R+T \eta_b)+1} -2\frac{ e^{-\frac{\eta_a|\alpha| ^2}{\eta_a \bar{n} R+1}}}{\eta_a \bar{n} R+1}-2\frac{ e^{-\frac{\eta_b|\beta| ^2}{\eta_b\bar{n}T+1}}}{\eta_b\bar{n}T+1}\Bigg].
\end{align}
\end{widetext}
The previous expression can easily be obtained by writing the thermal state as a mixture of coherent states $\rho_\text{th}( \bar n)= \frac{1}{\pi \bar n}\int e^{-\frac{|\alpha|^2}{\bar n}} \prjct{\alpha}d^2\alpha$, as the expectation value of W on a coherent state $\bra{\alpha} W \ket{\alpha}$ is easily obtained through the formula \eqref{Pns} using  $\bra{\alpha}(1-\eta)^{a^\dag a}\ket{\alpha}=  e^{-\eta |\alpha|^2}$.

\end{document}